\documentclass{proc-l}

\usepackage{graphicx}
\usepackage{xcolor}

\usepackage{verbatim}
\usepackage{amssymb}
\usepackage{mathtools}
\usepackage{tikz}
\usepackage{aliascnt}
\usepackage{hyperref}
\usepackage{cleveref}
\crefname{section}{Section}{Sections}
\crefname{equation}{equation}{equations}
\Crefname{equation}{Equation}{Equations}
\crefname{theorem}{Theorem}{Theorems}
\Crefname{theorem}{Theorem}{Theorems}
\crefname{proposition}{Proposition}{Propositions}
\Crefname{proposition}{Proposition}{Propositions}
\crefname{corollary}{Corollary}{Corollaries}
\Crefname{corollary}{Corollary}{Corollaries}
\crefname{lemma}{Lemma}{Lemmas}
\Crefname{lemma}{Lemma}{Lemmas}
\crefname{definition}{Definition}{Definitions}
\Crefname{definition}{Definition}{Definitions}
\crefname{example}{Example}{Examples}
\Crefname{example}{Example}{Examples}
\crefname{remark}{Remark}{Remarks}
\Crefname{remark}{Remark}{Remarks}
\crefformat{equation}{#2(#1)#3}
\Crefformat{equation}{#2(#1)#3}

\newtheorem{theorem}{Theorem}
\newaliascnt{proposition}{theorem}
\newtheorem{proposition}[proposition]{Proposition}
\aliascntresetthe{proposition}
\newaliascnt{corollary}{theorem}

\aliascntresetthe{corollary}
\newaliascnt{lemma}{theorem}
\newtheorem{lemma}[lemma]{Lemma}
\aliascntresetthe{lemma}

\theoremstyle{definition}
\newaliascnt{definition}{theorem}

\aliascntresetthe{definition}
\newaliascnt{example}{theorem}

\aliascntresetthe{example}

\theoremstyle{remark}
\newaliascnt{remark}{theorem}
\newtheorem{remark}[remark]{Remark}
\aliascntresetthe{remark}

\newcommand{\Z}{\mathbb{Z}}
\newcommand{\C}{\mathbb{C}}
\newcommand{\T}{\mathbb{T}}
\newcommand{\cH}{\mathcal{H}}
\newcommand{\ess}{\sigma_{\operatorname{ess}}}
\newcommand{\ran}{\operatorname{ran}}

\newcommand{\textbi}[1]{\textbf{\textit{#1}}}

\begin{document}

\title[Block encoding of the Segawa-Suzuki spectral mapping theorem]
{A block-encoding approach to the Segawa-Suzuki spectral mapping theorem}

\author{Yohei Tanaka}
\address{Department of Mathematical Sciences, College of Science and Engineering,
Ritsumeikan University,
1-1-1 Noji-higashi, Kusatsu, Shiga 525-8577, Japan}
\email{tana0035@fc.ritsumei.ac.jp}

\date{}
\dedicatory{}
\commby{}

\begin{abstract}
We revisit the Segawa--Suzuki spectral mapping theorem for abstract Szegedy-type quantum walks from a block-encoding viewpoint. Retaining both eigenspaces of the involution underlying the chiral symmetry yields a natural two-by-two block representation in which the usual discriminant operator and its counterpart form the diagonal blocks. The non-endpoint spectral mapping theorem then follows from an elementary Schur decomposition, substantially simplifying the original proof by avoiding its invariant-subspace decomposition and approximate-eigenvector arguments. The same framework establishes the result for the spectrum, point spectrum, and Fredholm essential spectrum and gives a direct block-theoretic description of the endpoint inherited and birth eigenspaces. We illustrate the construction with a one-dimensional split-step quantum walk.
\end{abstract}

\maketitle

\section{Introduction}

\begin{figure}[ht]
\centering
\begin{tikzpicture}[scale=1]
  \coordinate (O) at (0,0);
  \coordinate (Z) at (1.2,1.132);
  \coordinate (M) at (1.2,0);
  \coordinate (Zc) at (1.2,-1.132);
  \coordinate (ContL) at (-1.35,0);
  \coordinate (ContR) at (0.5,0);
  \coordinate (ContLP) at (145:1.65);
  \coordinate (ContLM) at (-145:1.65);
  \coordinate (ContRP) at (72.4:1.65);
  \coordinate (ContRM) at (-72.4:1.65);
  \draw[black,line width=2pt,line cap=butt] (72.4:1.65) arc[start angle=72.4,end angle=145,radius=1.65];
  \draw[black,line width=2pt,line cap=butt] (-72.4:1.65) arc[start angle=-72.4,end angle=-145,radius=1.65];
  \draw[->] (-2.2,0) -- (2.45,0) node[right] {\(\mathbf{Re}\)};
  \draw[->] (0,-2.2) -- (0,2.35) node[above] {\(\mathbf{Im}\)};
  \draw (O) circle (1.65);
  \draw[black,line width=2pt,line cap=butt] (ContL) -- (ContR);
  \draw[black,line width=1pt,line cap=butt] (ContL) -- (ContR);
  \fill (Z) circle (1.5pt) node[above right] {\(x+i\sqrt{1-x^2}\)};
  \draw[fill=white] (M) circle (1.5pt) node[above left] {\(x\)};
  \fill (Zc) circle (1.5pt) node[below right] {\(x-i\sqrt{1-x^2}\)};
  \draw[->,densely dotted, shorten <=3pt,shorten >=2pt] (M) -- (Z);
  \draw[->,densely dotted,shorten <=3pt,shorten >=2pt] (M) -- (Zc);
  \draw[->,densely dotted,shorten <=3pt,shorten >=2pt] (ContL) -- (ContLP);
  \draw[->,densely dotted,shorten <=3pt,shorten >=2pt] (ContL) -- (ContLM);
  \draw[->,densely dotted,shorten <=3pt,shorten >=2pt] (ContR) -- (ContRP);
  \draw[->,densely dotted,shorten <=3pt,shorten >=2pt] (ContR) -- (ContRM);
\end{tikzpicture}
\caption{The spectral correspondence \(x\mapsto x\pm i\sqrt{1-x^2}\) from \((-1,1)\) to \(\T\setminus\{-1,1\}.\) Since \(\|X_1\|\leq 1,\) its spectrum lies in \([-1,1],\) precisely the interval mapped to the unit circle by this correspondence.} 
\label{figure: spectral mapping}
\end{figure}
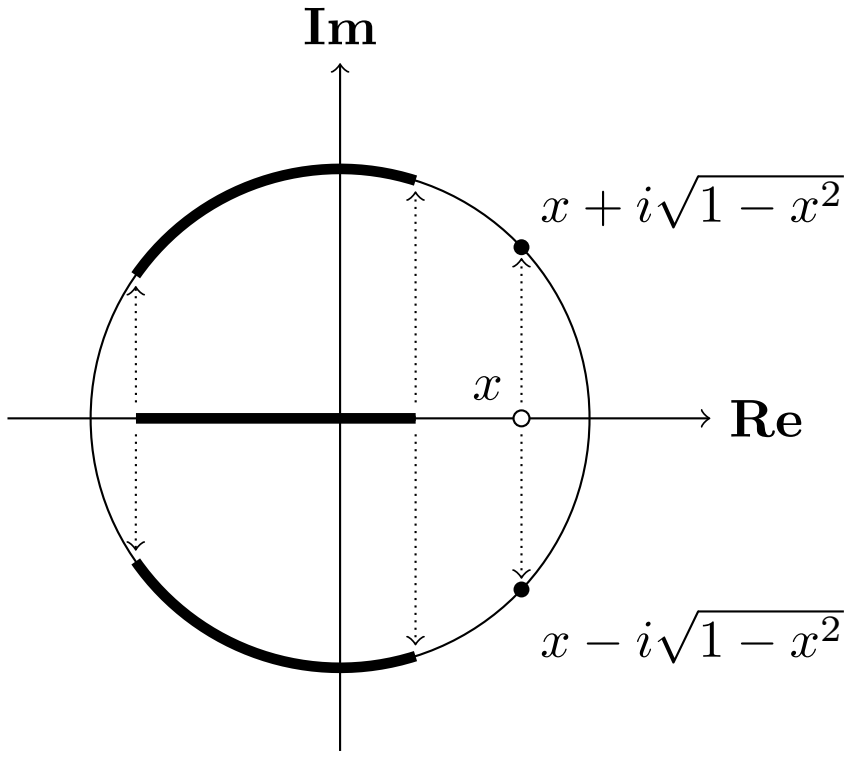

The \textbi{spectral mapping theorem} for quantum walks goes back at least to Szegedy's formulation \cite{Szegedy2004} and has since been developed in several forms, for example in \cite{HiguchiKonnoSatoSegawa2014,MatsueOgurisuSegawa2017,SegawaSuzuki2019}. Among them, the Segawa--Suzuki formulation in \cite{SegawaSuzuki2019} is the most abstract; for its historical background and further references, see the Introduction of that paper. In this formulation, one considers a unitary operator \(W_0\) on a Hilbert space \(\cH\) of the form
\begin{equation}
\label{equation: abstract Szegedy}
W_0 = S(2d_1^*d_1 - 1),
\end{equation}
where \(S\) is a unitary self-adjoint operator on \(\cH,\) and where \(d_1\) is a \textit{co-isometry} from \(\cH\) into an auxiliary Hilbert space \(\cH_1\) (i.e., \(d_1d_1^* = 1_{\cH_1}\)). This theorem associates with \(W_0\) a self-adjoint contraction \(X_1 := d_1 S d_1^*,\) called the \textbi{discriminant operator}, and asserts that, for each \(x \in (-1,1)\) and each \(y \in \{\sqrt{1-x^2},-\sqrt{1-x^2}\},\)
\begin{equation}
\label{equation: non-endpoint smt}
x \in \sigma_{\mathrel{\#}}(X_1) \mbox{ if and only if } x+iy \in \sigma_{\mathrel{\#}}(W_0),
\end{equation}
where the symbol \(\sigma_{\mathrel{\#}}\) denotes either the spectrum \(\sigma\) or the point spectrum \(\sigma_{\operatorname{p}}.\) Thus each point \(x\in(-1,1)\) on the real line corresponds to the two points \(x\pm i\sqrt{1-x^2}\) on the unit circle. It is this correspondence, illustrated in \cref{figure: spectral mapping}, that gives rise to the term \textbi{spectral mapping}.

In fact, the key structure underlying this spectral mapping is \textit{chiral symmetry;} we say that a unitary operator \(W\) on \(\cH\) is \textbi{chiral-symmetric} if 
\begin{equation}
\label{equation: chiral symmetry}
W^* = \varGamma W\varGamma
\end{equation}
for some unitary self-adjoint operator \(\varGamma\) on \(\cH.\) To see that the operator \(W_0\) defined by \cref{equation: abstract Szegedy} is chiral-symmetric, we set \(C:=2d_1^*d_1-1.\) The operator \(W_0\) can now be written as \(W_0 = SC,\) where \(S,C\) are unitary self-adjoint, and so \cref{equation: chiral symmetry} holds true with respect to \((\varGamma, W) = (S, W_0)\) and \((\varGamma,W) = (C, W_0).\) Consequently, \(\sigma_{\mathrel{\#}}(W_0)\) is symmetric with respect to the real axis, accounting for the two branches in the spectral mapping \cref{equation: non-endpoint smt} illustrated in \cref{figure: spectral mapping}. To pass from chiral symmetry to block encoding, we focus on the choice \((\varGamma,W)=(C,W_0)\) and use the standard spectral decomposition of \(C:\)
\[
C = (+1)d_1^*d_1 + (-1) (1-d_1^*d_1).
\]
Here, \(d_1^*d_1=(1+C)/2\) and \(1-d_1^*d_1=(1-C)/2\) are the spectral projections of \(C\) onto \(\ker(C-1)\) and \(\ker(C+1),\) respectively. The auxiliary Hilbert space \(\cH_1\) can be identified with the \(+1\) eigenspace of \(C\) via the isometry \(d_1^*,\) whereas an analogous identification of the \(-1\) eigenspace does not seem to be used explicitly in the existing literature. Our starting observation is that the two eigenspaces can be retained on an equal footing. Indeed, we can choose another co-isometry \(d_2:\cH\to\cH_2\) with \(d_2^*d_2=(1-C)/2,\) and show that, under the unitary identification \(\eta:=(d_1^*\quad d_2^*):\cH_1\oplus\cH_2\to\cH,\) the unitarily transformed operator \(W:=\eta^*W_0\eta\) has the following block representation (see \cref{proposition: block encoding of the discriminant operator}):
\begin{equation}
\label{equation: standard representation of W}
W =
\begin{pmatrix}
d_1Sd_1^* & -d_1Sd_2^* \\
d_2Sd_1^* & -d_2Sd_2^*
\end{pmatrix}_{\cH_1 \oplus \cH_2} =: 
\begin{pmatrix}
X_{1} & i Y_{2} \\
i Y_{1} & X_{2} \\
\end{pmatrix}_{\cH_1 \oplus \cH_2}.
\end{equation}
Here the usual discriminant \(X_1=d_1Sd_1^*\) is exactly the first diagonal block, while its natural counterpart, \(X_2=-d_2Sd_2^*,\) is the second diagonal block. The advantage of this block viewpoint is that Schur decomposition gives a significantly simplified proof of \cref{equation: non-endpoint smt} and simultaneously shows that the equivalence remains valid with \(X_1\) replaced by \(X_2.\) Similarly, at the endpoints \(x=\pm1,\) the same block viewpoint yields 
\begin{equation}
\label{equation: endpoint spectral mapping}
x \in \sigma_{\mathrel{\#}}(W_0) \mbox{ if and only if } x \in \sigma_{\mathrel{\#}}(X_1) \cup \sigma_{\mathrel{\#}}(X_2).
\end{equation}
For the particularly important case \(\sharp=\operatorname{p},\) the endpoint spectral mapping \cref{equation: endpoint spectral mapping} can be sharpened to the following well-known orthogonal decomposition with respect to \(\ker(C-1)\oplus\ker(C+1):\)
\[
\ker(W_0 \mp 1) = d_1^*\bigl(\ker(X_1 \mp 1)\bigr) \oplus \bigl(\ker(S \pm 1) \cap \ker d_1\bigr),
\] 
where the first and second summands are called the \textbi{inherited eigenspaces} and the \textbi{birth eigenspaces}, respectively \cite{MatsueOgurisuSegawa2017,FudaFunakawaSuzuki2018,Suzuki2019,KubotaSaitoYoshie2022}. We give this decomposition a new block-theoretic interpretation by showing that
\begin{align}
\label{equation1: birth inherited identities}
d_1^*\bigl(\ker(X_1 \mp 1)\bigr) &= \ker(S \mp 1) \cap \ker(C - 1) = \ker(S \mp 1) \cap \ker d_2, \\
\label{equation2: birth inherited identities}
d_2^*\bigl(\ker(X_2 \mp 1)\bigr) &= \ker(S \pm 1) \cap \ker(C + 1) = \ker(S \pm 1) \cap \ker d_1.
\end{align}
The identities above provide a unified treatment of the inherited and birth eigenspaces through the unitary identifications \(d_1^*\) and \(d_2^*:\) they are the endpoint eigenspaces of the diagonal blocks \(X_1 = d_1Sd_1^*\) and \(X_2 = -d_2Sd_2^*,\) respectively.

The paper is organised as follows. In \cref{section: abstract chiral spectral mapping}, we study an abstract chiral block representation of the form \cref{equation: standard representation of W}, establish the corresponding spectral mapping theorem for the spectrum and point spectrum, and extend the theorem to the (Fredholm) essential spectrum; we also give the endpoint decomposition and the unitary equivalence of the non-endpoint eigenspaces. In \cref{section: block encoding}, we return to \(W_0=SC,\) identify the discriminant and its counterpart with the two diagonal blocks, and recover the Segawa--Suzuki spectral mapping theorem. Finally, \cref{section: supplementary results} records further consequences and applies the construction to a one-dimensional Suzuki-type split-step quantum walk.

On a final note the novelty of our approach is particularly clear when compared with the original proof in \cite{SegawaSuzuki2019}, which proceeds through a detailed decomposition into invariant subspaces, followed by separate arguments for the point spectrum and the full spectrum, the latter involving approximate eigenvector sequences. By contrast, our approach avoids these technical steps altogether. By retaining both eigenspaces of \(C = d_1^*d_1 - d_2^*d_2,\) we obtain a natural \(2 \times 2\) chiral block representation in which the usual discriminant and its counterpart appear as the diagonal entries. The non-endpoint spectral mapping theorem then follows directly from an elementary Schur decomposition, giving a unified treatment of the spectrum, point spectrum, and essential spectrum. The same block representation also gives a direct description of the endpoint eigenspaces, placing the inherited and birth states on an equal footing.

\section{Main result}
\label{section: main result}
Let \(\cH_1\) and \(\cH_2\) be Hilbert spaces, and let \(W\) be a unitary operator on their orthogonal sum \(\cH_1\oplus\cH_2.\) Following the terminology of \cite{MatsuzawaSekiTanaka2021}, we call \((\cH_1,\cH_2,W)\) a \textbi{chiral triple} if \(W\) admits a block representation of the form
\begin{equation}
\label{Equation: Standard Representation}
W =
\begin{pmatrix}
X_1 & i Y_2 \\
i Y_1 & X_2 \\
\end{pmatrix}_{\cH_1 \oplus \cH_2},
\end{equation}
where we assume that \(X_1,X_2\) are self-adjoint, and that \(Y_2=Y_1^*.\) This is equivalent to saying that \cref{equation: chiral symmetry} holds true with respect to \(\varGamma := 1 \oplus -1\) on \(\cH_1 \oplus \cH_2.\) Given \cref{Equation: Standard Representation}, which we call the \textbi{chiral block representation} of \((\cH_1,\cH_2,W),\) the unitarity of \(W\) gives the following Pythagorean identities and the intertwining relations (see, for example, \cite[\textsection 2]{MatsuzawaSekiTanaka2021}):
\begin{enumerate}
\item We have \(X_j^2+Y_j^*Y_j = 1_{\cH_j}\) for each \(j=1,2.\)
\item We have \(X_1 Y_2 = Y_2 X_2\) (or, equivalently, \(X_2 Y_1 = Y_1 X_1\)).
\end{enumerate}

\subsection{Abstract chiral spectral mapping}
\label{section: abstract chiral spectral mapping}
Throughout this subsection, we assume that \((\cH_1,\cH_2,W)\) is a chiral triple and use its chiral block representation \cref{Equation: Standard Representation}.

\begin{theorem}[Abstract chiral spectral mapping for spectra]
\label{theorem2: abstract chiral spectral mapping theorem}
If the symbol \(\sigma_{\mathrel{\#}}\) denotes any one of the spectrum \(\sigma,\) the point spectrum \(\sigma_{\operatorname{p}},\) and the essential spectrum \(\ess,\) then for each \(x \in [-1,1]\) we have:
\begin{enumerate}
\item \textbf{Endpoint spectral mapping.} If \(|x| = 1,\) then 
\begin{equation*}
x \in \sigma_{\mathrel{\#}}(X_1) \cup \sigma_{\mathrel{\#}}(X_2) \mbox{ if and only if } x \in \sigma_{\mathrel{\#}}(W).
\end{equation*}
\item \textbf{Non-endpoint spectral mapping.} If \(|x|\neq 1,\) then the following equivalence holds true for each \(y \in \{\sqrt{1-x^2}, -\sqrt{1-x^2}\}:\)
\begin{equation*}
x+iy \in \sigma_{\mathrel{\#}}(W) \mbox{ if and only if } x \in \sigma_{\mathrel{\#}}(X_1) \mbox{ if and only if } x \in \sigma_{\mathrel{\#}}(X_2).
\end{equation*}
\end{enumerate}
\end{theorem}

Here, for a bounded operator \(A,\) we write \(\ess(A):=\{\lambda\in\C\mid A-\lambda \text{ is not Fredholm}\}.\)

\begin{proof}
Let \(\sigma_{\mathrel{\#}}\) denote \(\sigma,\) \(\sigma_{\operatorname{p}},\) or \(\ess.\)
(i) We shall make use of the general fact that given a bounded normal operator \(N,\) we have that \(N\) is invertible (resp.\ injective, Fredholm) if and only if \(N^*N\) is invertible (resp.\ injective, Fredholm). We let \(N := W \mp 1,\) and 
\[
(W\mp1)^*(W\mp1) =2(1\mp X) = 2(1 \mp X_1) \oplus 2(1 \mp X_2),
\]
where the last equality follows from \(X = X_1 \oplus X_2.\) It follows that \(W\mp1\) is invertible (resp.\ injective, Fredholm) if and only if each \(X_j \mp 1\) is invertible (resp.\ injective, Fredholm). The claim follows.

(ii) We set \(w := x+iy.\) Let us first express \(W-w\) as the following two Schur decompositions:
\begin{align*}
W-w
&= \begin{pmatrix}1&iY_2(X_2-w)^{-1}\\0&1\end{pmatrix}\begin{pmatrix}S_1(w)&0\\0&X_2-w\end{pmatrix}\begin{pmatrix}1&0\\i(X_2-w)^{-1}Y_1&1\end{pmatrix} \\
&=\begin{pmatrix}1&0\\iY_1(X_1-w)^{-1}&1\end{pmatrix}\begin{pmatrix}X_1-w&0\\0&S_2(w)\end{pmatrix}\begin{pmatrix}1&i(X_1-w)^{-1}Y_2\\0&1\end{pmatrix},
\end{align*}
where the operators \(S_1(w), S_2(w)\) are given by 
\begin{align*}
S_1(w) &:= X_1-w+Y_2(X_2-w)^{-1}Y_1 = X_1-w+(X_1-w)^{-1}Y_1^*Y_1, \\
S_2(w) &:= X_2-w+Y_1(X_1-w)^{-1}Y_2 = X_2-w+(X_2-w)^{-1}Y_2^*Y_2.
\end{align*}
Since \(w\in\T\setminus\{\pm1\},\) the operators \(X_j-w\) are invertible for \(j=1,2.\) Moreover, 
\begin{align*}
S_j(w) 
&:= X_j-w+(X_j-w)^{-1}(1 - X_j^2) \\
&= (X_j-w)^{-1}((X_j-w)^2+1-X_j^2) \\
&= -2w(X_j-w)^{-1}(X_j-x),
\end{align*}
where the last equality follows from \(2wx = w^2 + 1.\) Hence \(S_j(w)\) is invertible (resp.\ injective, Fredholm) if and only if \(X_j-x\) is invertible (resp.\ injective, Fredholm). The two Schur decompositions above therefore show that \(W-w\) is invertible (resp.\ injective, Fredholm) if and only if each \(X_j - x\) is invertible (resp.\ injective, Fredholm).
\end{proof}

\begin{remark}
As a byproduct of the same Schur decomposition, one also obtains an explicit resolvent formula; this will be recorded in \cref{proposition: resolvent formula}.
\end{remark}

For \(\sharp=\operatorname{p},\) the preceding spectral mapping only detects the presence of eigenvalues. The next theorem lifts this correspondence to the level of eigenspaces and hence also captures their multiplicities.

\begin{theorem}[Abstract chiral spectral mapping for eigenvalues]
\label{theorem: abstract chiral spectral mapping theorem}
The following assertions hold true for each \(x \in [-1,1]:\)
\begin{enumerate}
\item \textbf{Endpoint eigenvalue mapping.} If \(|x| = 1,\) then
\begin{equation}
\label{equation: endpoint SMT}
\ker(W - x) = \ker(X_1 -x ) \oplus \ker(X_2 - x).
\end{equation}

\item \textbf{Non-endpoint eigenvalue mapping.} If \(|x| \neq 1,\) then we have the following two unitary isomorphisms for each \(y \in \{\sqrt{1-x^2}, -\sqrt{1-x^2}\}:\)
\begin{align}
\label{equation: isomorphism}
&\ker(X_1 - x) 
\ni \psi_1 \longmapsto 
\frac{1}{\sqrt{2y^2}}
\begin{pmatrix}
y \psi_1 \\
Y_1\psi_1
\end{pmatrix}
\in \ker(W - (x+iy)), \\
\label{equation2: isomorphism}
&\ker(X_2 - x) 
\ni \psi_2 \longmapsto 
\frac{1}{\sqrt{2y^2}}
\begin{pmatrix}
Y_2\psi_2 \\
y \psi_2
\end{pmatrix}
\in \ker(W - (x+iy)).
\end{align}
\end{enumerate}
\end{theorem}
As shown in the proof below, the common eigenspace \(\ker(W - (x+iy))\) on the right-hand sides of \crefrange{equation: isomorphism}{equation2: isomorphism} admits the following description:
\begin{equation} 
\label{equation: eigenspace of W}
\ker(W - (x+iy)) = 
\left\{ 
(\psi_1, \psi_2)\in 
\begin{matrix}
\ker(X_1 - x) \oplus \ker(X_2 - x)
\end{matrix}
\mid Y_1 \psi_1 = y \psi_2 
\right\}.
\end{equation}
\begin{proof}
(i) This follows from \((W \mp 1)^*(W \mp 1) = \mp 2 (X \mp 1),\) where \(X = X_1 \oplus X_2.\)

(ii) We set \(w := x + iy.\) To prove \cref{equation: eigenspace of W} we first note that the left-hand side is a subspace of \(\ker(X - x) = \ker(X_1 - x) \oplus \ker(X_2 - x)\) by the polynomial functional calculus. It therefore suffices to show that given a vector $\psi$ of the form $\psi = \psi_1 \oplus \psi_2,$ where $\psi_j \in \ker(X_j - x)$ for each $j = 1,2,$ we have $\psi \in \ker(W - w)$ if and only if $Y_1 \psi_1 = y \psi_2.$ Note that we have 
\begin{align*}
(W - w)\psi 
=
\begin{pmatrix}
(X_1-w)\psi_1 + iY_2 \psi_2 \\
iY_1 \psi_1 + (X_2-w) \psi_2 
\end{pmatrix} 
=
\begin{pmatrix}
i(-y\psi_1 + Y_2 \psi_2) \\
i(Y_1 \psi_1 -y\psi_2) 
\end{pmatrix}.
\end{align*}
On the one hand, if $\psi \in \ker(W - w),$ then $iY_1 \psi_1 -iy \psi_2 = 0.$ On the other hand, if $Y_1 \psi_1 = y \psi_2,$ then $Y_2 \psi_2 = y \psi_1.$ We get $(W - w)\psi = 0.$ It follows from \cref{equation: eigenspace of W} that the mappings \cref{equation: isomorphism} and \cref{equation2: isomorphism} are well-defined isometries, since \(Y_j^* Y_j \psi_j = (1 - X_j^2)\psi_j = y^2 \psi_j\) for each \(j=1,2.\) The inverses are given respectively by 
\begin{align}
& \ker(W - w)
\ni 
\begin{pmatrix}
\phi_1 \\
\phi_2
\end{pmatrix}
\longmapsto 
\frac{\sqrt{2y^2}}{y} \phi_1 \in \ker(X_1 - x), \\
& \ker(W - w)
\ni 
\begin{pmatrix}
\phi_1 \\
\phi_2
\end{pmatrix}
\longmapsto 
\frac{\sqrt{2y^2}}{y} \phi_2 \in \ker(X_2 - x).
\end{align}
\end{proof}

\subsection{A block encoding of the discriminant operator}
\label{section: block encoding}

We now return to the setting of \cref{equation: abstract Szegedy}.

\begin{theorem}[Block encoding of the discriminant operator]
\label{proposition: block encoding of the discriminant operator}
Let \(\cH_1, \cH_2\) be two Hilbert spaces, and let \(S, C\) be two unitary self-adjoint operators on a Hilbert space \(\cH.\) For each \(j=1,2,\) let \(d_j : \cH \to \cH_j\) be an operator satisfying
\begin{equation}
\label{equation: decomposition of the abstract coin operator}
d_j d_j^* = 1_{\cH_j}, \qquad d_j^*d_j = \frac{1 +(-1)^{j+1} C}{2}.
\end{equation}
Then the following assertions hold true:
\begin{enumerate}
\item The \(1 \times 2\) block-operator matrix \(\eta := (d_1^* \quad d_2^*):\cH_1 \oplus \cH_2 \to \cH\) is unitary, and the operator \(C\) is diagonalisable as \(\eta^* C \eta = 1 \oplus -1\) with respect to \(\cH_1 \oplus \cH_2.\)  
\item If we set \((W,\varGamma) := (\eta^* SC \eta, \eta^* C \eta),\) then \((\cH_1,\cH_2,W)\) is a chiral triple whose chiral block representation is given by \cref{equation: standard representation of W}.
\item We have \(\ker(SC \mp 1) = d_1^*\bigl(\ker(X_1 \mp 1)\bigr) \oplus d_2^*\bigl(\ker(X_2 \mp 1)\bigr)\) with respect to \(\ker(C-1) \oplus \ker(C+1).\) Moreover, these two summands are identified explicitly in \crefrange{equation1: birth inherited identities}{equation2: birth inherited identities}.
\end{enumerate}
\end{theorem}
On a side note, the first equality in \cref{equation1: birth inherited identities} was proved in \cite[Corollary 3.2]{Suzuki2019}. The novelty of (iii) lies in treating \(d_1\) and \(d_2\) simultaneously, thereby identifying the inherited and birth eigenspaces symmetrically as the endpoint eigenspaces \(\ker(X_j \mp 1).\) 
\begin{proof}[Proof of \cref{proposition: block encoding of the discriminant operator}]
(i) 
\begin{align*}
\eta\eta^*
&= 
\begin{pmatrix}
d_1^* & d_2^*
\end{pmatrix}
\begin{pmatrix}
d_1 \\ d_2
\end{pmatrix}
=
d_1^*d_1+d_2^*d_2, \\
\eta^*\eta
&=
\begin{pmatrix}
d_1 \\ d_2
\end{pmatrix}
\begin{pmatrix}
d_1^* & d_2^*
\end{pmatrix}
=
\begin{pmatrix}
d_1d_1^* & d_1d_2^* \\
d_2d_1^* & d_2d_2^*
\end{pmatrix}
=
\begin{pmatrix}
1_{\cH_1} & d_1d_2^* \\
d_2d_1^* & 1_{\cH_2}
\end{pmatrix}.
\end{align*}
It follows from the first identity that \(\eta\eta^* = (1+C)/2 + (1-C)/2 = 1.\) On the other hand, \(\eta^*\eta = 1\) follows from 
\[
d_1d_2^* = d_1(\eta \eta^*) d_2^* = d_1 (d_1^*d_1+d_2^*d_2) d_2^* = 2d_1d_2^*.
\]
Moreover,
\[
\eta^*C\eta
=
\begin{pmatrix}
d_1 \\ d_2
\end{pmatrix}
C
\begin{pmatrix}
d_1^* & d_2^*
\end{pmatrix} 
=
\begin{pmatrix}
d_1 \\ d_2
\end{pmatrix}
\begin{pmatrix}
d_1^* & -d_2^*
\end{pmatrix} 
=
\begin{pmatrix}
1_{\cH_1} & 0 \\
0 & -1_{\cH_2}
\end{pmatrix},
\]
where the third equality follows from \((-1)^{j+1} C d_j^* = (2d_j^*d_j - 1)d_j^* = d_j^*\) for each \(j=1,2.\)

(ii) By the definitions of \(W\) and \(\varGamma,\)
\[
W^*
=\eta^*CS\eta
=\varGamma W\varGamma.
\]
Moreover, the same block multiplication gives
\[
W
=\eta^*SC\eta 
=
\begin{pmatrix}
d_1 \\ d_2
\end{pmatrix}
S
\begin{pmatrix}
d_1^* & -d_2^*
\end{pmatrix} 
=
\begin{pmatrix}
d_1Sd_1^* & -d_1Sd_2^* \\
d_2Sd_1^* & -d_2Sd_2^*
\end{pmatrix}_{\cH_1\oplus\cH_2}.
\]

(iii) By \cref{equation: endpoint SMT} applied to \(W=\eta^*SC\eta,\) and then conjugating by \(\eta,\) we obtain
\[
\ker(SC\mp1)
=d_1^*\bigl(\ker(X_1\mp1)\bigr)
\oplus d_2^*\bigl(\ker(X_2\mp1)\bigr).
\]
Let \(j,j'\in\{1,2\}\) with \(j\ne j'.\) Since \(d_jd_j^*=1\) and \(\ran d_j^*=\ran(d_j^*d_j)=\ker(C-(-1)^{j+1})=\ker d_{j'},\) the map \(d_j^*\) is unitary from \(\cH_j\) onto this subspace. Moreover, for \(f\in\cH_j,\)
\[
\|(S\mp1)d_j^*f\|^2
=2\langle f,(1\mp d_jSd_j^*)f\rangle,
\]
because \(S=S^*=S^{-1}.\) Hence \(d_jSd_j^*f=\pm f\) if and only if \(Sd_j^*f=\pm d_j^*f,\) and therefore
\[
d_j^*\ker(d_jSd_j^*\mp1)
=\ker(S\mp1)\cap\ker(C-(-1)^{j+1})
=\ker(S\mp1)\cap\ker d_{j'}.
\]
Thus these are the joint eigenspaces of the two involutions \(S\) and \(C,\) on which \(SC\) acts as \(\pm(-1)^{j+1}.\) This proves both identities.
\end{proof}

\begin{remark}
We show below that the usual block representation of \(SC\) induced by the eigenspace decomposition of \(C\) is recovered as a special case of \cref{proposition: block encoding of the discriminant operator}. Indeed, if we choose
\[
\cH_1 := \ker(C-1),
\qquad
\cH_2:=\ker(C+1),
\]
and define each \(d_j\) by \(\cH \ni \psi \mapsto (1 + (-1)^{j+1}C)/2 \psi \in \cH_j,\) then \(d_j^*\) becomes the canonical inclusion \(\cH_j\hookrightarrow\cH.\) It follows that \(\eta\) is the canonical unitary identification of \(\cH_1\oplus\cH_2\) with \(\cH.\) Consequently, \(W=\eta^*SC\eta\) is precisely the block representation of \(SC\) with respect to \(\ker(C-1)\oplus\ker(C+1) \cong \cH.\)
\end{remark}

\section{Supplementary results}
\label{section: supplementary results}

\subsection{Additional consequences of the block calculation}
\begin{proposition}[Unitary equivalence of the two discriminant eigenspaces]
With the notation introduced in \cref{theorem: abstract chiral spectral mapping theorem}, the following correspondence is unitary for each \(x\in(-1,1)\) and each \(y\in\{\sqrt{1-x^2},-\sqrt{1-x^2}\}:\)
\[
\ker(X_1 - x) \ni \psi \longmapsto y^{-1}Y_1 \psi \in \ker(X_2 - x).
\]
\end{proposition}
\begin{proof}
This follows by composing the two unitary isomorphisms in \cref{equation: isomorphism} and \cref{equation2: isomorphism}.
\end{proof}

\begin{proposition}[Resolvent formula]
\label{proposition: resolvent formula}
With the notation introduced in \cref{theorem2: abstract chiral spectral mapping theorem}, let \(x\in\rho(X_1)\cap\rho(X_2)\cap(-1,1),\) and let \(y\in\{\sqrt{1-x^2},-\sqrt{1-x^2}\}.\) Then \(w:=x+iy\) is in the resolvent set \(\rho(W),\) and
\begin{equation}
\label{equation: resolvent formula for W and X}
(W-w)^{-1}
= \frac{1}{2w}
\begin{pmatrix}
(X_1-x)^{-1}(w-X_1) & i(X_1-x)^{-1}Y_2 \\
i(X_2-x)^{-1}Y_1 & (X_2-x)^{-1}(w-X_2)
\end{pmatrix}.
\end{equation}
\end{proposition}
\begin{proof}
Either Schur decomposition used in the proof of \cref{theorem2: abstract chiral spectral mapping theorem} yields the same formula. Using the first one, we have
\[
W-w
= \begin{pmatrix}1&iY_2(X_2-w)^{-1}\\0&1\end{pmatrix}\begin{pmatrix}S_1(w)&0\\0&X_2-w\end{pmatrix}\begin{pmatrix}1&0\\i(X_2-w)^{-1}Y_1&1\end{pmatrix}.
\]
Since \(x \in \rho(X_1) \cap \rho(X_2),\) the proof of \cref{theorem2: abstract chiral spectral mapping theorem} gives
\[
S_1(w)^{-1}=\frac{1}{2w}(X_1-x)^{-1}(w-X_1).
\]
Therefore, \((W-w)^{-1}\) is given by 
\[
\begin{pmatrix}1&0\\-i(X_2-w)^{-1}Y_1&1\end{pmatrix}
\begin{pmatrix}S_1(w)^{-1}&0\\0&(X_2-w)^{-1}\end{pmatrix} 
\begin{pmatrix}1&-iY_2(X_2-w)^{-1}\\0&1\end{pmatrix}.
\]
Multiplying these factors and using \((w-X_1)Y_2=Y_2(w-X_2)\) and \(Y_1(w-X_1)=(w-X_2)Y_1,\) we obtain \cref{equation: resolvent formula for W and X}.
\end{proof}

\subsection{One-dimensional split-step quantum walk}

We consider the state Hilbert space \(\ell^2(\Z, \C^2)\) of \(\C^2\)-valued sequences defined on the one-dimensional integer lattice \(\Z.\) Up to unitary equivalence, the time-evolution operator \(W_0\) of the \textbi{one-dimensional (Suzuki-type) split-step quantum walk} is given by the following block-operator matrix on \(\ell^2(\Z, \C^2) = \ell^2(\Z,\C) \oplus \ell^2(\Z,\C):\)
\[
W_0 = \begin{pmatrix}
1 & 0 \\
0  &  L^*
\end{pmatrix}
\begin{pmatrix}
p & \sqrt{1-p^2} \\
\sqrt{1-p^2} & -p
\end{pmatrix}
\begin{pmatrix}
1 & 0 \\
0  &  L
\end{pmatrix}
\begin{pmatrix}
a & \sqrt{1-a^2} \\
\sqrt{1-a^2} & -a
\end{pmatrix},
\]
where \(L\) is the \textbi{left-shift operator} on \(\ell^2(\Z,\C),\) defined by \(L\psi := \psi(\cdot+1)\) for each \(\psi \in \ell^2(\Z,\C),\) and where the sequences \(p,a:\Z\to[-1,1]\) are viewed as bounded multiplication operators. This unitary operator \(W_0\) has the decomposition \(W_0=SC,\) where
\begin{align}
\label{equation: suzuki split-step quantum walk}
S
&:= 
\begin{pmatrix}
1 & 0 \\
0  &  L^*
\end{pmatrix}
\begin{pmatrix}
p & \sqrt{1-p^2} \\
\sqrt{1-p^2} & -p
\end{pmatrix}
\begin{pmatrix}
1 & 0 \\
0  &  L
\end{pmatrix}, \\
C &:=
\begin{pmatrix}
a & \sqrt{1-a^2} \\
\sqrt{1-a^2} & -a
\end{pmatrix}.
\end{align}

\begin{lemma}
\label{proposition: discriminant operators for SSQW}
In the notation of \cref{proposition: block encoding of the discriminant operator}, we put
\(\cH=\ell^2(\Z,\C^2)\) and \(\cH_1=\cH_2=\ell^2(\Z,\C).\) Let \(a_\pm := \sqrt{(1 \pm a)/2}.\)
For each \(j=1,2,\) we define \(d_j:\ell^2(\Z,\C^2)\to\ell^2(\Z,\C)\) through
\[
\begin{pmatrix}
d_1\\
d_2
\end{pmatrix}
:=
\begin{pmatrix}
a_+ & a_-\\
a_- & -a_+
\end{pmatrix}.
\]
Then \cref{equation: decomposition of the abstract coin operator} holds true for each \(j=1,2,\) and
\begin{align*}
X_1
&=
a_-L^*\sqrt{1-p^2}a_+
+a_+\sqrt{1-p^2}La_-
+a_+^2p-a_-^2p(\cdot-1), \\
X_2
&=
a_+L^*\sqrt{1-p^2}a_-
+a_-\sqrt{1-p^2}La_+
-a_-^2p+a_+^2p(\cdot-1), \\
iY_1
&=
-a_+L^*\sqrt{1-p^2}a_+
+a_-\sqrt{1-p^2}La_-
+a_+a_-(p+p(\cdot-1)).
\end{align*}
\end{lemma}
\begin{proof}
For \(t \in [-1,1],\) set
\[
R(t) := 
\begin{pmatrix}
t & \sqrt{1 - t^2} \\
\sqrt{1 - t^2} & -t
\end{pmatrix}.
\]
It is then easy to show that \(R(t)\) is diagonalisable by \(R\left(\sqrt{\frac{1+t}{2}}\right);\)
\[
R\left(\sqrt{\frac{1+t}{2}}\right) R(t) R\left(\sqrt{\frac{1+t}{2}}\right)
= 
\begin{pmatrix}
1 & 0 \\
0 & -1
\end{pmatrix},
\]
where
\[
R\left(\sqrt{\frac{1+t}{2}}\right) 
=
\begin{pmatrix}
\sqrt{\frac{1+t}{2}} & \sqrt{\frac{1-t}{2}} \\
\sqrt{\frac{1-t}{2}} & -\sqrt{\frac{1+t}{2}}
\end{pmatrix}.
\]
We have \(C = R(a)\) by using the functional calculus, and
\[
\eta = (d_1^* \quad d_2^*)
=
\begin{pmatrix}
a_+ & a_- \\
a_- & -a_+
\end{pmatrix}.
\]
Since \(\eta=\eta^*,\) the identity \(\eta^*C\eta=1\oplus(-1)\) gives
\[
C=\eta(1\oplus -1)\eta
= (d_1^* \quad d_2^*)
\begin{pmatrix}
1 & 0 \\
0 & -1
\end{pmatrix}
\begin{pmatrix}
d_1 \\ d_2
\end{pmatrix}
=d_1^*d_1-d_2^*d_2.
\]
The unitarity of \(\eta\) gives the two matrix identities
\[
\eta^*\eta
=
\begin{pmatrix}
d_1 \\ d_2
\end{pmatrix}
\begin{pmatrix}
d_1^* & d_2^*
\end{pmatrix}
=
\begin{pmatrix}
d_1d_1^* & d_1d_2^* \\
d_2d_1^* & d_2d_2^*
\end{pmatrix}
=
\begin{pmatrix}
1 & 0 \\
0 & 1
\end{pmatrix},
\]
and
\[
\eta\eta^*
=
\begin{pmatrix}
d_1^* & d_2^*
\end{pmatrix}
\begin{pmatrix}
d_1 \\ d_2
\end{pmatrix}
=d_1^*d_1+d_2^*d_2=1.
\]
Thus \(d_1d_1^*=d_2d_2^*=1,\) so each \(d_j\) is a co-isometry. Combining
\[
C=d_1^*d_1-d_2^*d_2,
\qquad
1=d_1^*d_1+d_2^*d_2
\]
gives \((-1)^{j+1} C=2d_j^*d_j-1\) for \(j=1,2.\)
It remains to compute the blocks explicitly. We have
\[
S=
\begin{pmatrix}
p & \sqrt{1-p^2}L \\
L^*\sqrt{1-p^2} & -L^*pL
\end{pmatrix}.
\]
Since \(L^*pL\) is the multiplication operator by \(p(\cdot-1),\) the identities
\(X_1=d_1Sd_1^*,\) \(X_2=-d_2Sd_2^*,\) and \(iY_1=d_2Sd_1^*\) give
\begin{align*}
X_1
&=
a_+pa_+
+a_+\sqrt{1-p^2}La_-
+a_-L^*\sqrt{1-p^2}a_+
-a_-L^*pLa_- \\
&=
a_-L^*\sqrt{1-p^2}a_++a_+\sqrt{1-p^2}La_-
+a_+^2p-a_-^2p(\cdot-1),
\end{align*}
and
\begin{align*}
X_2
&=
-a_-pa_-
+a_-\sqrt{1-p^2}La_+
+a_+L^*\sqrt{1-p^2}a_-
+a_+L^*pLa_+ \\
&=
a_+L^*\sqrt{1-p^2}a_-+a_-\sqrt{1-p^2}La_+
-a_-^2p+a_+^2p(\cdot-1).
\end{align*}
Finally,
\begin{align*}
iY_1
&=
a_-pa_+
+a_-\sqrt{1-p^2}La_-
-a_+L^*\sqrt{1-p^2}a_+
+a_+L^*pLa_- \\
&=
-a_+L^*\sqrt{1-p^2}a_++a_-\sqrt{1-p^2}La_-
+a_+a_-(p+p(\cdot-1)).
\end{align*}
\end{proof}

In particular, the usual discriminant operator is the first diagonal block \(X_1=d_1Sd_1^*.\) When \(p\) is constant, the formula for \(X_1\) above recovers the expression obtained in \cite[Lemma 3.3]{FudaFunakawaSuzuki2018}.

\bibliographystyle{amsalpha}
\bibliography{references}

\end{document}